# Optimization of High Entropy Alloy Catalyst for Ammonia Decomposition and Ammonia Synthesis

*Wissam A. Saidi[1]\** *, Waseem Shadid[2], and Götz Veser[3]*

[1]*Department of Mechanical Engineering and Materials Science, University of Pittsburgh, Pittsburgh, PA 15261, USA*

[2]*Department of System and Information Systems, The University of North Carolina Charlotte, Charlotte, North Carolina, USA*

[3]*Department of Chemical and Petroleum Engineering, University of Pittsburgh, Pittsburgh, PA 15261, USA*

*\*corresponding author: alsaidi@pitt.edu*

## Abstract

The successful synthesis of high entropy alloy (HEA) nanoparticles, a long-sought goal in materials science, opens a new frontier in materials science with applications across catalysis, structural alloys, and energetic materials. Recently, a $Co_{25}Mo_{45}Fe_{10}Ni_{10}Cu_{10}$ HEA made of earth-abundant elements was shown to have a high catalytic activity for ammonia decomposition, which rivals that of state-of-the-art, but prohibitively expensive, ruthenium catalyst. Using a computational approach based on first-principles calculations in conjunction with data analytics and machine learning, we build a model to rapidly compute the adsorption energy of H, N, and $NH_x$ ($x = 1,2,3$) species on CoMoFeNiCu alloy surfaces with varied alloy compositions and atomic arrangement. We show that the 25/45 Co/Mo ratio identified experimentally as the most active composition for ammonia decomposition increases the likelihood that the surface adsorbs nitrogen equivalently to that of ruthenium while at the same time interacting moderately strongly with intermediates. Our study underscores the importance of computational modeling and machine learning to identify and optimize HEA alloys across their near-infinite materials design space.

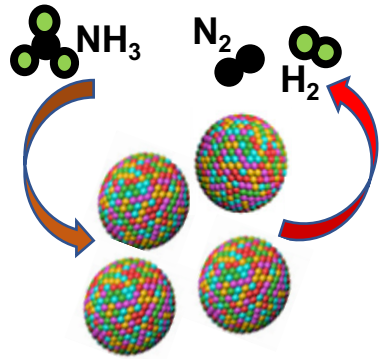

The recent successful synthesis of high-entropy alloy (HEA) nanoparticles paves the way to unexplored opportunities in materials science.[1-3] Initial findings in this nascent field strongly suggest that HEAs may in particular enable the design and synthesis of novel catalysts with higher activity, lower cost, or both.[4-8] For example, CrMnFeCoNi HEA NPs are demonstrated to have an intrinsic electrocatalytic activity for oxygen reduction which exceeds that of the precious metal Pt.[5] AuAgPtPdCu HEAs are shown to be efficient catalysts for the electrochemical reduction of $CO_2$.[6] However, the essentially unlimited number of HEA combinations calls for the development of a science-based approach to identify and engineer catalysts with desired activity, selectivity, and stability.

Recently, Wang and collaborators showed that HEAs made up of earth-abundant elements Co, Mo, Fe, Ni, and Cu can decompose ammonia 20 times more efficiently than the state-of-the-art precious-metal ruthenium catalysts.[9] By synthesizing 5 different alloys with similar concentrations for Fe, Ni, and Cu but varied Co and Mo compositions, they showed that $Co_{25}Mo_{45}Fe_{10}Ni_{10}Cu_{10}$ had the highest catalytic activity and confirmed from kinetic measurements that it has the lowest reaction barrier. Further, this study showed that the optimum composition binds nitrogen similar to ruthenium, while others have higher or lower nitrogen adsorption energies. The present study uses high-throughput screening in conjunction with data analytics based on machine learning (ML) to rationalize the experimental results, and offers a practical approach to identify further CoMoFeNiCu alloy candidates with high catalytic efficiency.

Ammonia is a strategic industrial chemical that, via fertilizers, is largely responsible for sustaining the growing global population. Currently, more than 146 million tons of ammonia are synthesized annually via the Haber-Bosch (HB) process, which operates at temperatures of 400 – 600 ºC and pressures between 100 – 300 bar to convert molecular hydrogen and nitrogen into ammonia: $3H_2(g) + N_2(g) \rightarrow 2NH_3$; $\Delta H = -0.96$ eV. Also, ammonia represents a promising sustainable liquid fuel as an energy carrier for mobile and remote applications[10].  However, existing catalysts for ammonia synthesis and ammonia decompositions have severe limitations. The industrial catalyst for ammonia synthesis, based on the century-old discovery that Fe enables $N_2$ activation, is no longer considered sustainable despite its wide use: Its demand for high purity hydrogen, obtained in practice from fossil fuels via steam reforming, results in ~2.5% of worldwide fossil fuel-based $CO_2$ emissions, equating to ~670 million tons per year.[11] Furthermore, the relatively low activity of the Fe-based industrial catalyst contributes to the very high energy demand of the HB process, which is estimated to consume 1-2% of global primary energy:  The low activity requires high reaction temperatures to achieve economical rates, which then demand very high operating pressure to overcome thermodynamic limitations at these temperatures to make the process yields economically feasible.[11-12] On the other hand, the most active known catalyst to date for ammonia synthesis and decomposition, the precious metal Ru, is not economical due to its prohibitive cost.[13]    Unsurprisingly, there is hence great interest in finding more active and cost-effective catalysts for both ammonia synthesis and decomposition.

Traditional approaches for catalyst design, which rely on an "Edisonian" trial-and-error approach, are slow, expensive, and likely to result in suboptimal catalysts. For instance, Bosch's technician, Alwin Mittasch, tested more than 3,000 catalyst compositions for ammonia synthesis in over 20,000 experiments at the beginning of the 20[th] century to optimize the Fe-based catalyst.[14]  In contrast, computational modeling based on quantum-mechanical density functional theory (DFT) provides a rational approach to catalyst design by uncovering the reaction mechanism and energetics at the molecular level.[15] Although ammonia decomposition operates at different reaction conditions than ammonia synthesis,[16] both reactions share the same similar mechanisms and *a volcano-type relationship between the catalytic activity and nitrogen adsorption energy*.[17-21]  Thus, using this simple descriptor, trends from one material to the next can be easily identified, which thus allows for the possibility of identifying new catalysts with near-optimal descriptor values.[22]

Scaling relations and the Brønsted-Evans-Polanyi (BEP) relationship which are widely used in catalysis, do not generally hold for alloy surfaces.[23] The standard descriptor-based screening approach will hence provide a necessary but not sufficient filtering.[24-26] However, a recent experimental study of ammonia decomposition unambiguously indicated that nitrogen adsorption energy is a good descriptor for the catalytic efficiency of CoMoFeNiCu HEAs.[9] Also, a recent study utilized binding energies as a descriptor to optimize AgAuCuPdPt HEAs for $CO_2$ and CO reduction[3], and results were validated experimentally[6]. Hence, we posit that a screening based on nitrogen adsorption energy is a viable approach to optimize HEA catalysts for ammonia decomposition and, by extension, to ammonia synthesis. This is also supported by the current findings that are found to be in excellent agreement with experimental results[9] for ammonia decomposition.

The stability of alloys as a single solid solution – rather than undesirable ordered intermetallics that dissociate into multiple phases – is a challenge for HEA synthesis. As the name HEA implies, it was believed that the high configurational entropy is essential for forming a single-phase solid solution; however, it is now generally appreciated that the definition can be broader as numerous so-called HEAs do not have high configurational entropies.[27] Yang and Zhang empirically introduced simple rules to assess the HEA stability similar to the classic Hume–Rothery rules that delineate the conditions under which an element can dissolve in metal to form a solid solution.[28] More specifically, it is empirically found that a $k$–element alloy will form a stable HEA phase if $\delta r < 6.6\%$ and $\Omega = (T_m \, \Delta S_{\text{mix}})/|\Delta H_{\text{mix}}| > 1.1$, where $\delta r$ is the difference in atomic radius, $T_m$ the melting temperature, $\Delta S_{\text{mix}}$ and $\Delta H_{\text{mix}}$ the mixing entropy and enthalpy, respectively, and weighted averages are defined in terms of the concentration ($c_i$) as $\delta r = \sum_{i,j}^k c_i c_j (r_i - r_j)$, $\Delta H_{\text{mix}} = \sum_{i,j}^k c_i c_j H_{ij}$, $\Delta S_{\text{mix}} = -k_B \sum_i^k c_i \ln c_i$, and $T_m = \sum_i^k c_i \, T_m^i$.[28-29] Not surprisingly, there have been exceptions to the Yang-Zhang rules, which have prompted several modifications such as accounting for non-rigidity of the radii or adding more features within an ML approach.[30-31]

To assess the HEA stability of CoMoFeNiCu alloys, we compute $\delta r$ and $\Omega$ values for $250,000$ different alloys. The set is constructed by varying the composition of the elements from 2% to 96% in 2% steps and choosing compositions that add up to 100%. All the proposed combinations are found to satisfy the empirical constraints for HEA stability except ~0.04% that are found deficient in more than two metal components. In agreement with these findings, the previous experimental study confirmed that the synthesized $Co_xMo_yFe_{10}Ni_{10}Cu_{10}$ HEAs $(x/y = 15/55, 25/45, 35/35, 45/25,$ and $55/15)$ have homogeneous mixing of the elements with a random atom-to-atom contrast variation based on STEM-EDS and atomistic simulations.[9] Thus, hereafter we assume that the alloys will form a single HEA phase.

Further, consistent with the experimental results, we assume that the alloy will adapt a single face-centered cubic (fcc) lattice. In our study, we calculate the lattice constant of the HEAs as a weighted average based on the alloy composition, following Vegard's law for binary alloys. For the same chemical composition that was synthesized experimentally, the HEA lattice constant varies between 3.60 Å for $Co_{55}Mo_{15}Fe_{10}Ni_{10}Cu_{10}$ and 3.73 Å for $Co_{15}Mo_{55}Fe_{10}Ni_{10}Cu_{10}$. The reported[9] experimental interplanar (111) spacing in the HEA is measured as 2.18 Å suggesting a 3.74 Å lattice constant consistent with our estimated values.

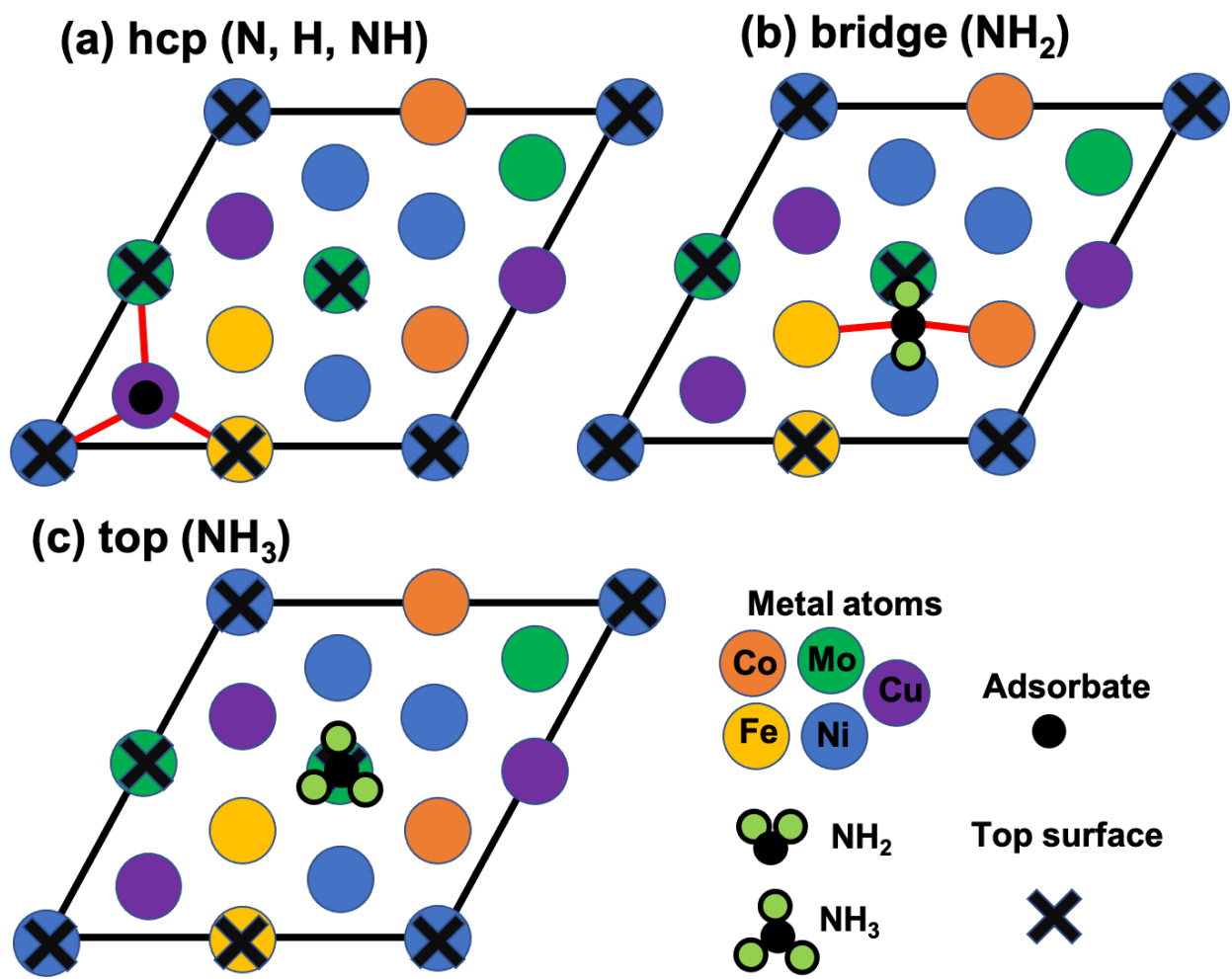

**Figure 1.** Top view for adsorption configurations for different chemical species on a representative HEA fcc(111) surface. Atoms belonging to the top surface layer are labeled with a cross. Bonds between adsorbates and metal atoms are shown as a red line.

We employ a slab approach within the (111) fcc termination to investigate the interactions of nitrogen with HEA surfaces. The slab model is a good approximation to describe the catalytic activity of the relatively large ~20 nm HEA nanoparticles synthesized experimentally, which were reported to have well-resolved (111) fcc planes.[9] However, given the exceeding large number of different alloy configurations (e.g. there are $3 \times 10^{11}$ configurations for realizing a (2x2x5) supercell of an equiatomic alloy), it is prohibitive to evaluate nitrogen adsorption energies on all sites to statistically determine the activity using DFT calculations. Instead, we follow an approach recently demonstrated by Rossmeisl and coworkers who developed an ML approach to model the adsorption energy $\Delta E_{ML}$ chemical species on HEA surfaces. $\Delta E_{ML}$ assumes a linear dependence on the number of metal elements in the chemical environment of the adsorption site as $\Delta E_{ML} = \sum_z \sum_k C_{z,k} N_{z,k}$. Here $N_{z,k}$ is the number of metal elements of type $k$ in zone $z$ and $C_{z,k}$ are corresponding fitting parameters. The zones are designed based on geometrical distances with respect to the adsorption site. Despite its simplicity, this ML model showed good agreement with DFT values.[3, 32]

Herein we follow a similar approach to build a ML model for nitrogen adsorption energy but with important modifications that appreciably improve its fidelity and transferability. First, we utilize a convolutional neural network (CNN) that can capture nonlinear correlations in the training set, offering higher flexibility compared to the previous linear model[3, 32]. Second, we designed robust features that encode three levels of domain knowledge: (1) element-specific features comprising of ionization energy, electronegativity, electron affinity, number of valence electrons; (2) metal-specific features including Wigner-Seitz radius $r_s$, d-band center $\epsilon_d$, d-band filling $f_d$, coupling matrix elements between adsorbate and metal d-states $V_{ad}$, $d\ln\epsilon_d/d\ln r_s$, and workfunction; and (3) chemical environment descriptors corresponding to the identity of metals neighboring the adsorption site. Thus, each site will yield a different adsorption energy given the wide variations of the associated features. In our geometric neighbor analysis, we account for the maximum information available in the surface microstructure by including all metal atoms based on geometrical distance from the adsorption site. The feature set was found to saturate by including only the top 3 surface layers. Clearly, this representation is equivalent to that obtained from the spherical zone representation[32] except that it alleviates the repetitiveness of some features since some atoms are equivalent due to periodic boundary conditions. Additionally, in our feature description, we include the overall composition of the alloy, which we also find to further improve the ML model fidelity.

We perform a high-throughput DFT computation for nitrogen adsorption on various compositions of CoMoFeNiCu HEA (111) surfaces. We find that nitrogen adsorbs most strongly at hexagonal-closed pack (hcp) sites and hence we focused our ML on this adsorption configuration (see Figure 1). In our database, we have a total of 1,911 configurations that differ in the alloy composition and arrangement of the metal atoms. As can be seen from the histogram in Figure S1, nitrogen binding energies in the database vary between -2.4 and 1.2 eV. The large variance in the adsorption energies is not surprising given the large number of different chemical environments associated with the adsorption site. However, while the wide variations ensure that the training dataset captures a wide range of different systems, this increases the required training time for accurate ML. Further to take advantage of symmetry, we extended the dataset by 10-25% by including configurations related to an original configuration by a permutation of selected surface atoms. For instance, for the hcp site, the adsorption energies associated with different arrangements of the 3 nearest-neighbor elements of the adsorption site are nearly the same. We have used 80% of the dataset selected at random to train the CNN model, and equally split the remaining dataset for validation and testing. Cross-validation of the results is performed on 5 different models based on a different selection of the training dataset to quantify the uncertainty of predictions.

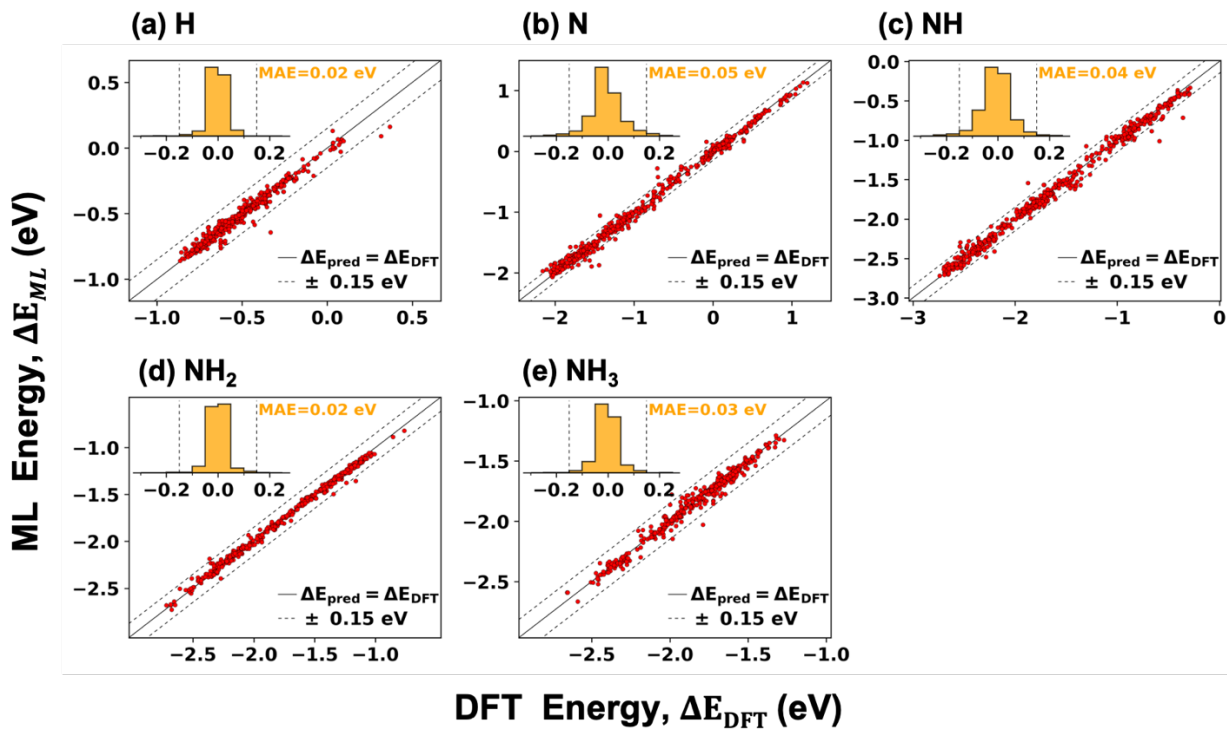

**Figure 2.** Comparison between adsorption energy predictions based on ML and DFT ground truth values on the testing set for (a) N, (b) H, (c) NH, (d) NH$_2$, and (e) NH$_3$. The inset shows the distribution in the differences between the ML and DFT values. The mean absolute error (MAE) shows the overall accuracy of the model. All energies are in eV.

Figure 2 compares the DFT calculated adsorption energies and CNN model predictions. As seen from the figure, the accuracy of the CNN model is notably high, achieving a mean-absolute error (MAE) of 0.05 eV/N (per nitrogen atom). Almost all of the ML predictions are within 0.15 eV from the DFT reference values. Importantly, we find no systematic bias in these predictions, which makes the accuracy of these calculations comparable to the intrinsic accuracy of the DFT approach employed to generate the training dataset.

Using the CNN model and our selected features, we evaluate nitrogen binding energy on HEA (111) surfaces. We focused on Co$_x$Mo$_y$Fe$_{10}$Ni$_{10}$Cu$_{10}$ HEA compositions to compare with experimental results where $x + y = 70$. For each alloy composition defined by $x$ and $y$, we compute the average adsorption energy $\hat{E}^{*,N} = \sum_{\ell}^{n_s} f_\ell E_\ell^{*,N} / \sum_{\ell}^{n_s} f_\ell$ over a large number of slab models generated randomly with different atomic arrangements. Here $E_\ell^{*,N}$ is the ML adsorption energy for $\ell$th configuration from the $n_s$ ensemble – we find that $n_s = 800$ yields converged adsorption energies (see Figure S2). Given the molar fractions $c_k$ of the HEA, $f = \prod_k^5 c_k^{N_k}$ accounts for the different possibilities of generating a surface microstructure with $N_k$ metal atoms (here we dropped the configuration label $\ell$ for clarity).[3, 32]

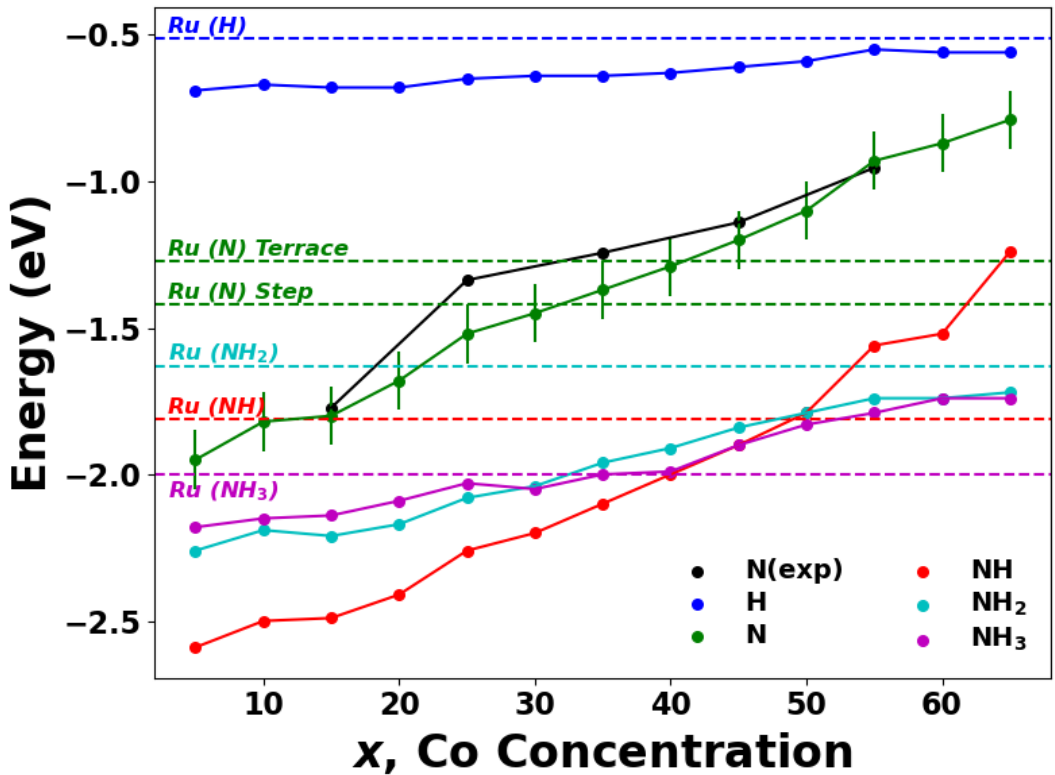

**Figure 3.** Adsorption energies of H, N, NH, NH$_2$, and NH$_3$ on Co$_x$Mo$_y$Fe$_{10}$Ni$_{10}$Cu$_{10}$ (111) as a function of Co concentration (with $x + y = 70$). The dashed horizontal lines (color-coded) show the corresponding adsorption energy values on Ru (0001) terraces. Also, we show the N adsorption energy for Ru (0001) on a step-edge. For nitrogen adsorption energy, we show statistical error based on ensemble averaging and ML uncertainty. Similar errors are expected for other adsorption energies and are omitted for clarity. The experimental results for nitrogen adsorption energy (shifted by 0.5 eV) are also shown for comparison.

Figure 3 shows the resulting $\hat{E}^{*,N}$ for $Co_xMo_yFe_{10}Ni_{10}Cu_{10}$ as a function of the Co concentration, $x$. As seen from the figure, the Co-rich HEA surface has weak adsorption energy while the Mo-rich surface binds nitrogen more strongly. These trends are inherited from the pure metals: Co surface binds nitrogen too weakly while Mo surface binds nitrogen too strongly. To establish limiting values, we find that nitrogen adsorption energies on Co and Mo fcc (111) are -0.56 and -2.51 eV/N, respectively. Thus, the HEA interpolates between the two limits. The concept of interpolation between strong and weak adsorption sites is not new: this was the basis for suggesting CoMo as a potential catalyst for ammonia synthesis more than twenty years ago[18]. Indeed, this prediction was verified recently in an experimental study showing that CoMo binary alloy with uniform distribution of the elements is a good catalyst for ammonia synthesis[33].

We also compare in Figure 3 the computed nitrogen binding energy with the experimental results[9] for the 5 different compositions. To mitigate the strong binding tendency of the PBE functional[34], we shifted all experimental values by 0.5 eV. Experimentally, it was determined that the optimum ammonia decomposition catalyst binds nitrogen with a strength similar to that of Ru (0001).[9] As seen from the figure, there is an excellent correlation between the CNN-computed and experimental nitrogen binding energies.

Importantly, the average adsorption energy $\hat{E}^{*,N}$ in Figure 3 is shown to vary linearly with Co concentration $x$ (correlation coefficient, $r^2 = 0.99$). This suggests that we can readily optimize the alloy composition, i.e. $x$, to increase the likelihood that the surface microstructure will bind nitrogen optimally. For ammonia decomposition, Ru is the most efficient catalyst. For Ru (0001), we find that the optimum nitrogen adsorption energy is located at the short-bridge site and has binding energy $E_{Ru}^{*,N} = -1.27$ eV/N. This agrees with previous results.[35] Figure S3 compares the PBE and revised PBE functional that has been employed in previous studies [36]. Thus, from Figure 2, we conclude that HEAs with $x = 0.35 - 0.45$ bind nitrogen with a strength equivalent to that of Ru (0001) and are thus expected to be efficient catalysts for ammonia decomposition. Further, using the nitrogen binding energies on Fe surfaces as target values, we can extend these arguments for ammonia synthesis. For Fe (110) and Fe (111), we investigated different adsorption sites using the employed computational setup and found the optimum values to be -1.82 and -1.70 eV/N, respectively. This suggests that $Co_xMo_yFe_{10}Ni_{10}Cu_{10}$ HEAs with $x \sim 0$ that are mostly depleted of Co serve as good catalysts for ammonia synthesis.

Our estimated value of the optimum value of Co concentration for efficient ammonia decomposition $x = 0.35 - 0.45$ is overestimated compared to the experimentally determined value, $x = 0.25$. While such discrepancies between the modeling and experimental results are not uncommon, we posit that this discrepancy is also partly because we used nitrogen binding energy on Ru (0001) terraces as a target value. The experimental results are obtained on nanoparticles; hence step-edges rather than terraces should provide the target value for optimization. Further, it has previously been determined that the B5 sites are the most catalytically active.[37] The B5 site corresponds to an arrangement of three Ru atoms in one layer and two further Ru atoms in the layer directly above at a monoatomic step on a Ru(0001) terrace. Using the nitrogen binding energy on Ru step edge that is lower than the terrace value by 0.1 eV,[36] we can infer from Figure 3 that the predicted optimum alloy composition corresponds to $x = 0.25 - 0.35$, which is in good agreement with experimental results.[9]

The generally accepted mechanism for ammonia decomposition involves the stepwise dissociation of H atoms from the $NH_3$ molecule on the metal catalyst surface. The reverse hydrogenation steps comprise the prevalent process for ammonia synthesis from $H_2(g)$ and $N_2(g)$ based on the Haber-Bosch process. Thus, for both reactions, there are 4 chemical species $NH_3$, $NH_2$, NH, and H interacting with the surface in addition to nitrogen. To assess how these intermediates interact with HEA, we developed ML adsorption models for each species, as we did for nitrogen. The favorable adsorption site for each species is found by inspecting different sites for several alloy samples. We find that $NH_3$ binds the most strongly at the top metal sites as is in the case of Ru (0001)[36] due to bonding via the lone electron pair on the N atom. $NH_2$ is found to bind to a twofold bridge site, and NH is found to bind to the hcp site.

See Figure 1 for a schematic illustrating these bonding configurations. All of these results are consistent with previous findings.[36] The comparison between the ML predictions and DFT results in Figure 2 shows that the ML models achieve for all species similarly high fidelity as for nitrogen.

Figure 3 shows the average adsorption energies for all intermediates for $Co_xMo_yFe_{10}Ni_{10}Cu_{10}$ alloys.  For comparison, we also show in the figure (dashed lines) the corresponding adsorption values for Ru (0001). As Co concentration increases, we see a small but noticeable increase in the hydrogen binding energy. However, all HEA compositions are found to adsorb hydrogen more strongly than Ru (0001).  In contrast, $NH_x$ ($x$ = 1, 2, 3) adsorption energies strongly decrease as the alloy becomes richer in Mo, similar to the case of nitrogen. This behavior is also inherited from the parent metals Co and Mo.  Interestingly, as seen in the figure, the adsorption energies of $NH_x$ are correlated with that of nitrogen. For the pure metal surfaces, there is a strong linear relationship between the adsorption energy of $AH_x$ ($x$= 1, 2, 3) for A= C, N, S, O with that of the A atom on a range of different d-block transition metals surfaces.[21] Our results show that these scaling relationships persist for the HEA although are weaker than on metallic surfaces.  Further, for CoMoFeNiCu, we see a weaker correlation with the N adsorption energies for the cases of $NH_2$ and $NH_3$. This can be understood because of the differences in the chemical environment of the adsorption sites: N binds to an hcp site while $NH_2$ and $NH_3$ adapt bridge and top configurations, respectively. NH, on the other hand, that adapts an hcp site similar to that of N shows a stronger correlation in its adsorption energies with the corresponding nitrogen values.  We posit that these appreciably preserved correlations between $NH_x$ and N adsorption energies explain in part why nitrogen is a good descriptor for ammonia synthesis on these HEAs.

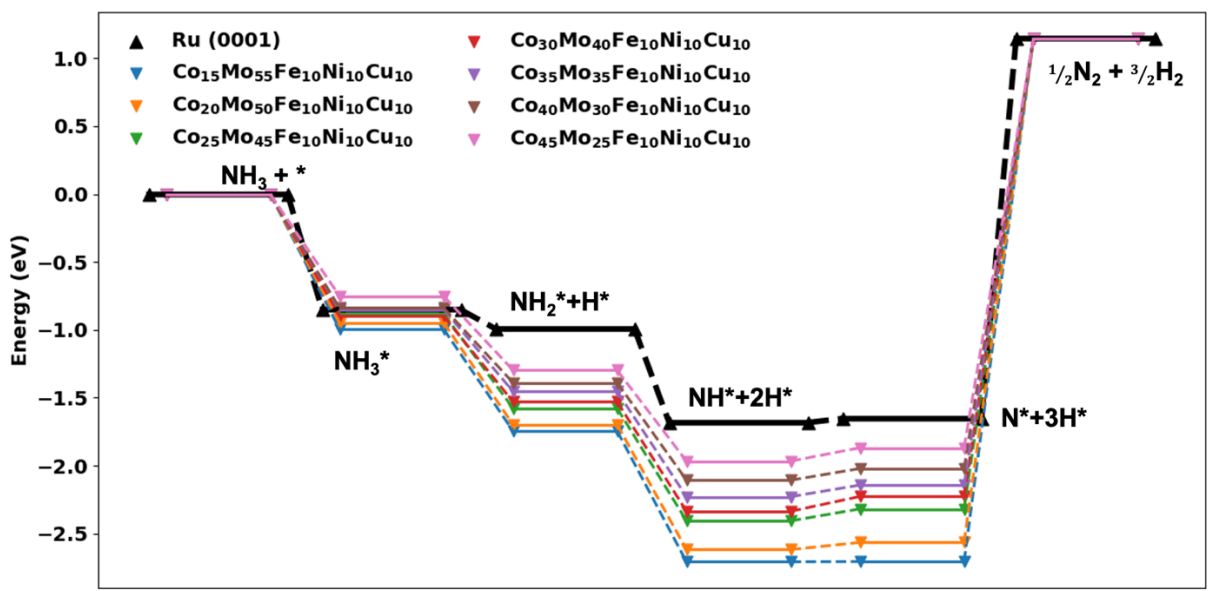

**Figure 4.** Ammonia decomposition energy landscape for $Co_xMo_yFe_{10}Ni_{10}Cu_{10}$ corresponding to different Co/Mo concentrations. The corresponding Ru (0001) values are shown for comparison. * corresponds to an empty adsorption site on the slab and A* indicates a site occupied by species A.

Using the computed adsorption energies for nitrogen and the intermediates, we can examine the full energy pathway during the dehydrogenation steps of $NH_3$ and compare it with the corresponding values for Ru (0001). As shown in Figure 4, the optimum catalysts adsorb $NH_3$ similarly to Ru (0001) while the Co-rich HEAs adsorb $NH_3$ weakly and Mo-rich ones adsorb $NH_3$ strongly. Contrastingly, the HEAs are found to bind the intermediates more strongly than Ru resulting in more exothermic decomposition reactions than Ru (0001). Interestingly, the highly catalytic step edges in Ru (0001) are also found to bind the intermediates more strongly than the terrace states.[36] However, a strong enhancement of the intermediates' bonding could lead to unfavorable surface poisoning by the intermediates and catalyst deactivation. From Figure 3, we see that in the optimum doping range, the $NH_2$ intermediate is stabilized appreciably more than the Ru (0001) case.

The experimental study examined a minute part of the composition space of the HEA with equal concentrations of Fe, Ni, and Cu, but it varied amounts for Co and Mo. Thus, it is not clear whether the full optimization of the alloy composition could result in an improved catalyst. Clearly, this problem cannot be solved experimentally given the infinite number of compositions that are expected to form a stable HEA phase as inferred from Yang-Zhang empirical rules. Using our developed computational approach, we can address this challenge and narrow down potential alloy compositions. To achieve this goal, we employ a particle swarm algorithm (PSO) for unbiased optimization by minimizing $|\hat{E}^{*,N} - E_{opt}^{*,N}|$ where $E_{opt}^{*,N} = 1.27$ for ammonia decomposition and 1.8 eV for ammonia synthesis. The PSO is performed by using a 256 population size and 20 generations. Also, we limited the search to alloys that have the 5 metal components present with a concentration larger than 1%. The PSO identified several potential HEAs for both reactions, including several HEAs for ammonia decomposition with significantly lower Co content than the previously identified HEA, such as $Co_{0.17}$ $Mo_{0.25}$ $Fe_{0.30}Cu_{0.18}$ $Ni_{0.10}$, which reduces both the Co and Mo content (the two by far most expensive elements in the set) by almost half while tripling the content in cheap and abundant Fe. For a complete list, see Tables S1 and S2

In conclusion, we have developed an approach based on machine learning and global optimization to optimize the activity of CoMoFeNiCu towards ammonia decomposition and ammonia synthesis. Particularly, optimizing the ratio of Co to Mo concentrations in $Co_xMo_yFe_{10}Ni_{10}Cu_{10}$ suggests that a ratio of $x{:}y$ ~25:45 – 35:35, i.e. slightly Mo-rich compositions, yields HEAs with a similar nitrogen binding energy to that of Ru (0001). These findings agree with recent experimental results, which identified $Co_{25}Mo_{45}Fe_{10}Ni_{10}Cu_{10}$ as a highly efficient ammonia decomposition catalyst. Further, we show that the scaling relationships between the binding energies of $NH_x$ ($x = 1,2,3$) and N still hold in CoMoFeNiCu similar to monometallic surfaces. These correlations explain in part why nitrogen binding energy is a good descriptor for ammonia decomposition and, by extension, for ammonia synthesis on these HEAs. The developed methodology can be applied to the discovery and optimization of HEAs for other catalytic reactions.

**Methods.** The DFT calculations employ the Perdew-Burke-Ehrenzhof exchange-correlation functional and projector augmented wave (PAW) pseudopotentials[38-39] as implemented in the Vienna Ab initio Simulation Package (VASP) package. We expanded the electronic wavefunctions using plane-wave representation with a 300 eV cutoff. The slab fcc (111) models are represented using a 2x2x5 supercell approach. For Ru (0001) we used a 2x2x5 supercell while as for bcc Fe (110) and Fe (111) models, we used 2x2x5 and 2x2x8 supercells, respectively. These slab models are expected to have small finite-size effects based on previous investigations[36]. We sampled the Brillouin zone using a 3x3x1shifted Monkhorst-Pack grid with 0.2 eV Gaussian smearing. All of the atomic coordinates belonging to the top two layers are relaxed if all forces are less than 0.1eV/Å and the energy changes are less than $10^{-5}$ eV in the self-consistent electronic step. Further, all calculations are performed with spin-polarized orbitals.

The adsorption energy $\Delta E_X$ for chemical species X is calculated as,

$$\Delta E_X = E_X^* - E^* - E_{ref}$$

where $E_X^*$ is the energy of the relaxed slab with the adsorbed species, $E^*$ is the energy of the relaxed surface, and $E_{ref}$ is properly normalized energy measured with respect to $H_2$ and $N_2$. The lattice parameters of the slabs are obtained from the weighted-average DFT-calculated lattice parameters for atoms belonging to the top layer following Vegard's law. The approximated lattice constants are shown to be in good agreement with experimental values.

The CNN architecture follows our previous study.[40] Briefly, it consists of an input layer that is passed to two convolutional layers, followed by one fully connected layer, then an output layer. The input layer is the set of features of size 137 that encode the domain knowledge of the composition of interest. The first convolutional layer consists of 64 filters, employs an element-wise rectified linear activation, and uses a one-dimensional (1D) convolutional kernel. The size of this kernel is equal to the number of input features, which allows the neural network to capture relationships that incorporate all input features. The output of this layer for each filter has the same size as the input. Then a max-pooling layer is used to downsample the output by a factor of two. The second convolutional layer consists of 128 filters, employs an element-wise rectified linear activation, and uses a 2D convolutional kernel. The number of rows of this kernel is equal to five, while the number of columns is set to the number of input features divided by two. A 2D kernel is used in the second layer rather than a 1D kernel to allow the network to utilize the 2D output (filters and features) of the 1$^{st}$ layer. The output of this layer for each filter has the same size as the input divided by two due to the previous downsampling layer. Then another max-pooling layer is used to downsample the output by a factor of two. Downsampling is used to help in compressing the features between convolutional layers.[41] The fully connected layer is composed of 100 hidden neurons that are connected to the down-sampled output of the second convolutional layer. Each neuron applies an element-wise rectified linear activation. Then a dropout layer with a drop rate of 0.2 is used to prevent overfitting. The last layer consists of one neuron that takes the output of the fully connected layer as input and estimates the adsorption energy. This neuron applies an element-wise rectified linear activation. This deep convolutional network design has been implemented using TensorFlow Python API.[42]

**ACKNOWLEDGMENTS**

W. A. S. is grateful for the U. S. National Science Foundation (Award No. CSSI-2003808). Also, we are grateful for computing time provided in part by the Pittsburgh Center for Research Computing (CRC) resources at the University of Pittsburgh and Argonne Leadership Computing Facility, a DOE Office Science User Facility supported under Contract DE-AC02-06CH11357.

**Supplemental Material Statement.**  Histogram of training data set, convergence profiles, comparison between PBE and RPBE reaction energies,  list of potential catalysts obtained from full optimization of the HEA alloy for ammonia decomposition and ammonia synthesis.